\documentclass{jfm}

\usepackage{graphicx}
\usepackage{float}
\usepackage{newtxtext}
\usepackage{newtxmath}
\usepackage{bm}
\usepackage{natbib}
\usepackage{hyperref}
\hypersetup{
    colorlinks = true,
    urlcolor   = blue,
    citecolor  = black,
}
\usepackage[noabbrev]{cleveref}
\usepackage{multirow}
\usepackage[normalem]{ulem} 
    \let\isout\sout \renewcommand{\sout}[1]{\ifmmode\text{\isout{\ensuremath{#1}}}\else\isout{#1}\fi}

\newcommand{\RomanNumeralCaps}[1]
\linenumbers

\newcommand{\unit}[1]{\,\mathrm{#1}}

\title{Hydrodynamic loads and vortex evolution from a bio-inspired pectoral fin near a solid body}

\author{Xiaowei He\aff{1,2}
  \corresp{\email{xiaowei.he@utah.edu}}
 \and Kenneth Breuer\aff{1}}

\affiliation{\aff{1}Center for Fluid Mechanics, School of Engineering, Brown University, Providence, Rhode Island 02912, USA
\aff{2}Department of Mechanical Engineering, University of Utah, Salt Lake City, Utah 84112, USA}

\begin{document}
\maketitle

\begin{abstract}
    A fin-body configuration is tested in a water tunnel to study the hydrodynamic loads and vortex evolution under dynamic fin-flapping motions, which is an idealized approximation of the pectoral fins of fish.
    The fin flaps about its leading edge, which is attached to the side of the body, at a range of combinations of amplitudes ($0^\circ-30^\circ$) and frequencies ($0.25\unit{Hz}-2\unit{Hz}$ or $k=0.16-1.26$), so the Strouhal number ($St=0.013-0.419$).
    The quasi-steady hydrodynamic loads exhibit significant hysteresis during the upstroke and downstroke phases of the fin flapping.
    Particle image velocimetry (PIV) measurements show the details of the shear layer and vortex development in dynamic flapping cases.
    Orbiting behaviors of the fin tip vortices are observed in larger Strouhal number cases.
    PIV results also reveal the influence of vortices on hydrodynamic loads in terms of lift fluctuations and thrust generation.
    The strong dependency on the reduced frequency and Strouhal number leads to scalings of the hydrodynamic loads using a data-driven method to select highly correlated terms.
    The most significant terms selected by the scaling process are quadratic terms of the Strouhal number and its nonlinear combinations with the reduced frequency.
\end{abstract}



\section{Introduction}
\label{sec:intro}
The hydrodynamics of fish swimming has long been a focus of research, and with the rapid development of robotic underwater vehicles, it has received considerable attention in recent years. Biological studies on how fish utilize their bodies to interact with their fluid environment include characterization of individual body kinematics \citep{domenici1997,liao2003} to locomotion \citep{lauder2015,saadat2017} to schooling behavior \citep{hemelrijk2015,ligman2023}. From a fluid mechanical perspective, the locomotion of fish or fish-like swimmers has also focused on multiple perspectives, including idealized first principles \citep{lighthill1975:Mathematicalbiofluiddynamics},  traveling wave propagation \citep{videler1999,muller2002}, vortex dynamics and wakes \citep{drucker1999,videler1999,brouzet2025}, and the effects of added mass \citep{paniccia2021}.  

Most studies have focused on the propulsive aspects of fish swimming, and furthermore on the role of the tail, or caudal fin.  This is a large body of literature and several measurements, theories and simulations characterizing the forces, scaling, and efficiency as functions of fin motion, flexibility and geometry have been reported by numerous research groups, \citep[e.g.][]{triantafyllou2000:HydrodynamicsFishlikeSwimminga, bandyopadhyay2002, green2011:unsteadythreedimensionalwake, quinn2014heaving, dewey2013, floryan2017, vanBuren2019,floryan2019:Largeamplitudeoscillationsfoils,maertens2015,brouzet2025};  although there is much still to learn, this area is relatively well-characterized from both the biological and fluid mechanical perspectives.

In contrast, less work has been reported on the roles of other fin groups - notably the pectoral, or side fins - and on their role both in propulsion as well as maneuvering of fish and fish-inspired robots. The pectoral fins are known for their multi-functional roles, and fish use them in different ways to generate different kinds of forces (propulsive, maneuvering, braking, etc) \citep{walker1997}.  Measurements of live fish have reported on detailed kinematics of the three-dimensional pectoral fin motion \citep{blake1979:MechanicsLabriformLocomotion,walker1997,bozkurttas2009:Lowdimensionalmodelsperformance} and have visualized the flow structures generated by the fin motion \citep{drucker2002piv,drucker2002pectoralfunction,lauder2006:Locomotionflexiblepropulsors}. Measurements have inspired computational simulations \citep{beal2007, bozkurttas2009:Lowdimensionalmodelsperformance,dong2006,dong2010} and robotic models \citep{kato1998:Locomotionmechanicalpectoral,kato2000:Controlperformancehorizontal,suzuki2008:Loadcharacteristicsmechanical,tangorra2010} which yield forces and flow structures associated with both rigid and flexible pectoral fin motion. 

Nevertheless, while these studies have been foundational in establishing the performance of the pectoral fins for propulsion and maneuvering, they have some limitations. Firstly, while the forces and flow generated by the fully three-dimensional (and in some cases flexible) fin, some of the more fundamental flow mechanisms are obscured due to the complexity of the geometry and kinematics. Secondly, some studies characterize the fin-induced flows without considering the interaction with the fish body \citep{beal2007,dong2010}. Lastly, in measuring and modeling a specific fin geometry and motion, the generation of scaling laws for the flows and forces over a range of fin sizes, amplitudes, and frequencies of operation has not been possible, thus limiting the extension of the work to more general applications.

To address this, we have pursued an intermediate path and experimentally characterized the forces and flow structures generated by the motion of an \emph{idealized} pectoral fin moving in a sinusoidal flapping motion with the pivot located at a rigid wall (representing the fish body). In order to clearly identify canonical flow structures, the fin is represented by a two-dimensional rigid plate. Time-resolved force measurements (streamwise and transverse) are synchronized with velocity measurements (two-dimensional particle image velocimetry)
and the fin is operated over a range of flapping amplitudes, $A_\theta$, and frequencies, $f$, thus covering a broad range of non-dimensional Strouhal numbers, $St = fA_\theta/U_\infty$, and reduced frequencies, $k = \pi f c/U_\infty$ (here $U_\infty$ is the free stream velocity and $c$ a characteristic scaling length).  The scaling of these forces is achieved using a data-driven approach \citep{brunton2016} to help select high-correlation parameters that optimally represent the lift and drag generated by the fin motion.

\section{Experiments}
\label{sec:exp}
    \subsection{Experimental setup}
    \label{sec:exp:setup}
\begin{figure}
        \centering
        \includegraphics[width=0.9\textwidth]{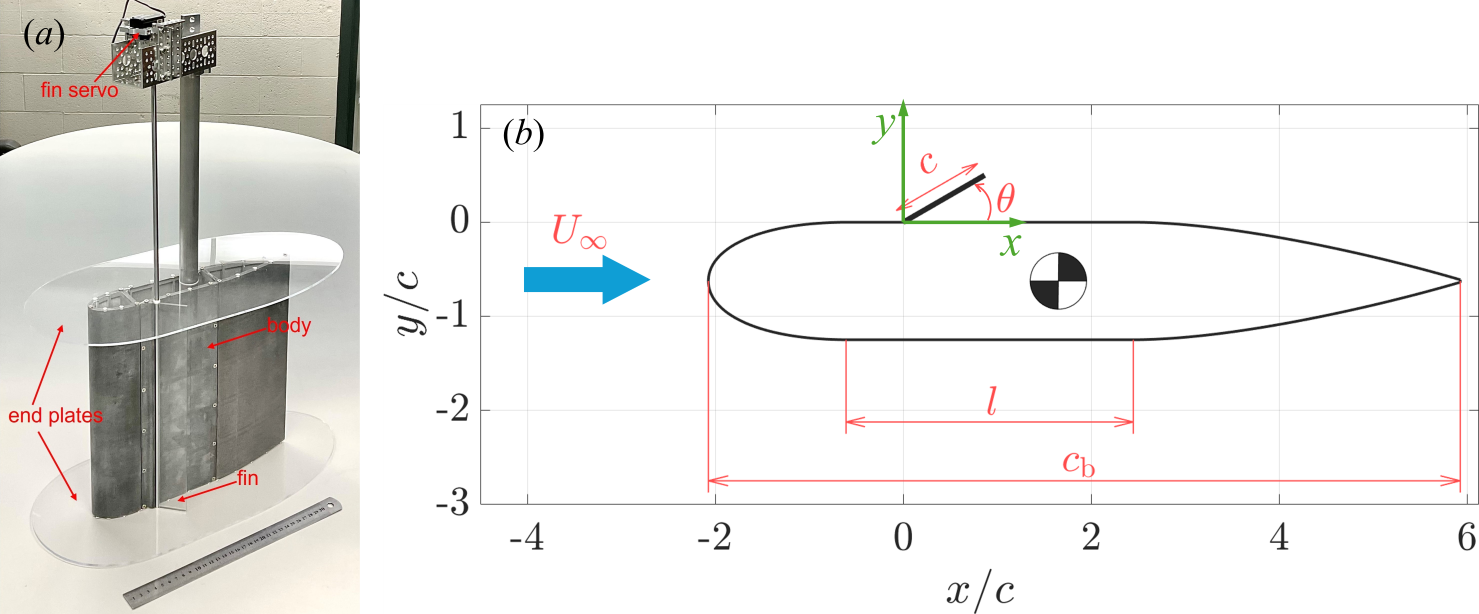}
        \caption{Fin-body experimental setup: (a) physical model; (b) coordinates and geometry normalized by fin chord length $c$.}
        \label{fig:exp:setup}
    \end{figure}

    The experiments were conducted in the free-surface water tunnel at Brown University.
    The test section of the water tunnel has a cross-section dimension of $W\times D = 80\unit{cm}\times 60\unit{cm}$.
    The physical fin-body model is shown in \cref{fig:exp:setup} and consists of a streamlined two-dimensional body 3-D printed with Nylon, with a straight two-dimensional fin mounted on one side.
    A pair of end plates is mounted on the model to prevent 3-D effects.
    The two-dimensional ``fish'' body has a cross-section of a modified NACA 0015 airfoil with a constant thickness section of length $122.2\unit{mm}$ inserted at the point of maximum thickness ($-0.16 \leqslant x/c \leqslant 2.45$) in \cref{fig:exp:setup}b).
    The overall dimensions of the fish are $320\unit{mm}$ by $50\unit{mm}$. The height of the model from one end plate to another is $30\unit{cm}$.
    
    The origin of the coordinate system is located at the fin leading edge (flapping hinge) with the $x$ axis pointing to the downstream direction and the $y$ axis pointing in the normal direction.
    The fin, made of clear acrylic, has a chord, $c$, of $40\unit{mm}$ and a thickness of $3.2\unit{mm}$ and is mounted $0.61c$ from the start of the constant thickness section.
    The fin pitches about its leading edge, controlled by a servo motor mounted above the waterline through a rigid shaft.
    The fin flapping angle, $\theta$, is measured with an optical encoder attached to the fin shaft with a $\pm 0.15^\circ$ uncertainty.
    A positive flapping angle is defined in the counterclockwise direction when the fin flaps away from the body.  For the results discussed in this manuscript, we consider periodic flapping with frequency $f$ and amplitude $A_\theta$ with a mean position such that the fin lies flat with the body ($\theta = 0$) at the bottom of the downstroke.
        
    The entire model is mounted on a six-axis force-torque transducer (ATI Gamma) through a rigid metal shaft with the $z$ axis of the force transducer aligned with the shaft axis, which is also the center of gravity location.
    The forces and torques were recorded at a sampling rate of $1\,000\unit{samples/s}$.
    A $10\unit{Hz}$ low-pass filter was applied in force data processing (5-40 times the flapping frequency).
    The force measurements are decoupled into lift, $L$, in the normal direction aligned with the body $y$ axis and drag, $D$, in the streamwise direction aligned with the body $x$ axis.
    Although the changes in the hydrodynamic loads are generated due to the fin position and motion, the external forces act on and are measured with respect to the entire fin-body configuration.
    For this reason, we choose to use the body chord length, $c_\mathrm{b}$, and the body platform area, $S$, to define the hydrodynamic load coefficients---lift coefficient, $C_L$, and drag coefficient, $C_D$:
    \begin{equation}
        \displaystyle
        C_L = \frac{L}{\frac{1}{2}\rho U_\infty^2 S}\quad \text{and}\quad 
        C_D  = \frac{D}{\frac{1}{2}\rho U_\infty^2 S}.
    \end{equation}
    
    We fixed the freestream velocity at $U_\infty=0.2\unit{m/s}$ for all cases in this study.
    Thus, the fin chord referenced Reynolds number is $Re_\mathrm{f}=7\,600$ and the body chord referenced Reynolds number is $Re_\mathrm{b}=60\,000$.
    
    Planar two-component particle image velocimetry (PIV) measurements are performed and synchronized with the force measurements.
    The PIV data are taken at a sampling rate of 100 image pairs per second.
    The PIV processing uses a $32\times32$-pixel interrogation window with a $50\%$ overlap, which results in a resolution of $1\unit{mm}$ in both $x$ and $y$ directions.Post-processing of the PIV data includes phase averaging over 30 cycles (for $k=0.157$ and $k=0.314$ cases), 60 cycles (for $k=0.628$ cases), or 120 cycles (for $k=1.257$ cases).
    Nondimensionalization of the flow field uses the fin chord length, $c$, as the length scale and the freestream velocity, $U_\infty$, as the velocity reference.
    The nondimensional $z$-vorticity is thus given by $\omega_z^* = \omega_zc/U_\infty$, and nondimensional $u$-velocity is given by $u^* = u/U_\infty$.
    
    After phase averaging, a complete flapping cycle consists of 60 evenly spaced phases (frames).
    Since the PIV measurements are synchronized with the hydrodynamic load measurements in each case, the same phase averaging is also applied to the hydrodynamic load data.
    
    \subsection{Test cases}
    \label{sec:exp:test}
    The unsteady flapping cases are organized by the sinusoidal motion of the fin, which varies in amplitude, $A_\theta$, and frequency, $f$:
    \begin{equation}
        \theta(t) = A_\theta(1 - \cos2\pi ft),
    \end{equation}
    where $A_\theta  = [7.5^\circ$, $15^\circ$, $30^\circ]$, and $f = [0.25, 0.5, 1, 2]$ Hz. The reduced frequency, $k$, is defined as
    \begin{equation}
        k = \frac{\pi fc}{U_\infty},
        \label{eq:exp:test:k}
    \end{equation}
    and the Strouhal number, $St$, is determined by the flapping frequency, the peak-to-peak arc length traveled by the fin tip, and the freestream speed: 
    \begin{equation}
        \displaystyle
        St = \frac{f(2A_\theta\pi/180^\circ)c}{U_\infty}.
        \label{eq:exp:test:St}
    \end{equation}
    The unsteady flapping experiments were conducted over twelve combinations of the flapping amplitude, $A_\theta$, and frequencies, $f$.
    These combinations and their corresponding Strouhal numbers are 
    summarized in \cref{tab:k-St}.
    \begin{table}
        \begin{center}
            \def~{\hphantom{0}}
            \setlength{\tabcolsep}{12pt}
            \begin{tabular}{c c c c c l}
                ~ & $f=0.25\unit{Hz}$ & $f=0.5\unit{Hz}$ & $f=1\unit{Hz}$ & $f=2\unit{Hz}$ & ~\\ \noalign{\smallskip}
                $k$ & $0.157$ & $0.314$ & $0.628$ & $1.257$ & ~\\ \noalign{\smallskip}
                $A_\theta=7.5^\circ$ & $0.013$ & $0.026$ & $0.052$ & $0.105$ & \multirow[c]{3}{*}{$\Biggl\}\;St$}\\
                $A_\theta=15^\circ$ & $0.026$ & $0.052$ & $0.105$ & $0.209$\\
                $A_\theta=30^\circ$ & $0.052$ & $0.105$ & $0.209$ & $0.419$
            \end{tabular}
            \caption{Reduced frequencies, $k$, and Strouhal numbers, $St$, of test cases.}
            \label{tab:k-St}
        \end{center}
    \end{table}
    
    In order to define a baseline reference, a quasi-steady flapping case is tested, where the fin flaps slowly from $0^\circ$ to $100^\circ$ and back to $0^\circ$ at a constant flapping rate of $2^\circ/\mathrm{s}$, which corresponds to $k=0.006$.

\section{Results and discussion}
\label{sec:results}
    \subsection{Baseline (quasi-steady) forces}
    \label{sec:results:baseline}
\begin{figure}
        \centering
        \includegraphics[width=0.8\textwidth]{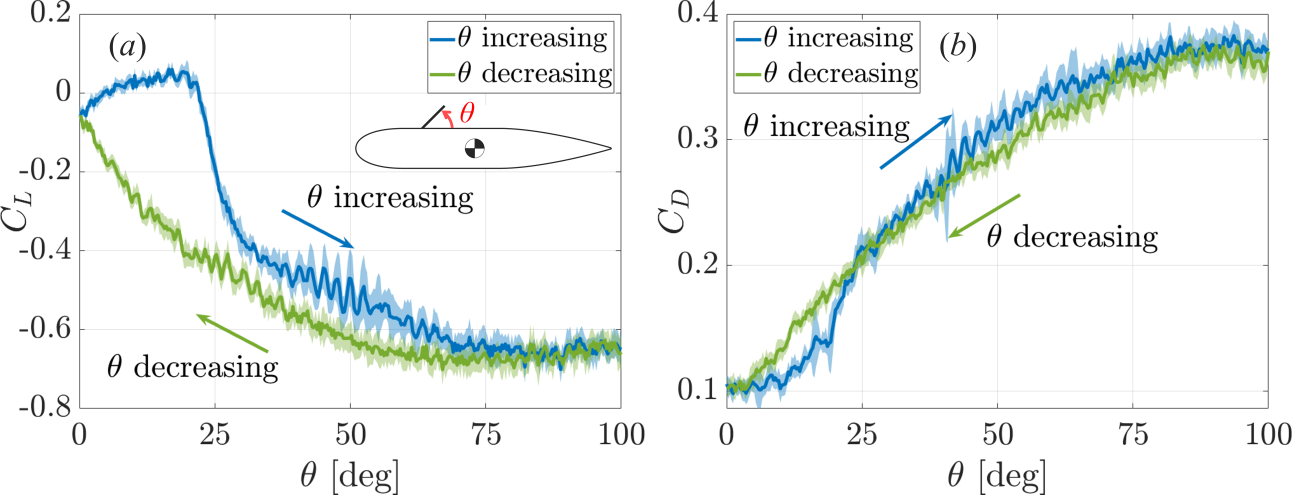}
        \caption{Baseline hydrodynamic loads vs. fin flap angle $\theta$: (a) lift coefficient $C_L$; (b) drag coefficient $C_D$; blue lines: upstroke ($\theta$ increasing) phase; green lines: downstroke ($\theta$ decreasing) phase; shades: standard deviation interval ($1\sigma$).}
        \label{fig:results:baseline}
    \end{figure}

    The baseline measurements from the quasi-steady tests are shown in \cref{fig:results:baseline}, presenting the lift (\cref{fig:results:baseline}a) and drag (\cref{fig:results:baseline}b) coefficients, plotted against the fin flapping angle.
    The flap-up ($\theta$ increasing) and flap-down ($\theta$ decreasing) phases of the fin flapping are marked by blue and green colors, respectively.
    The baseline lift coefficient varies from zero to $-0.7$ as the flapping angle changes.
    The maximum change in the absolute values occurs at $\theta\approx76^\circ$.
    We observe vibrations in the lift and drag coefficient baselines, which may be related to the natural shedding frequency of the fin tip vortices.
    The uncertainty indicated by the standard deviation intervals (shaded area in \cref{fig:results:baseline}) increases at higher fin angles ($\theta>35^\circ$) as the amplitudes of the vibrations become larger, which is a result of the more energetic vortex shedding at these fin angles.
        
\begin{figure}
        \centering
        \includegraphics[width=0.8\textwidth]{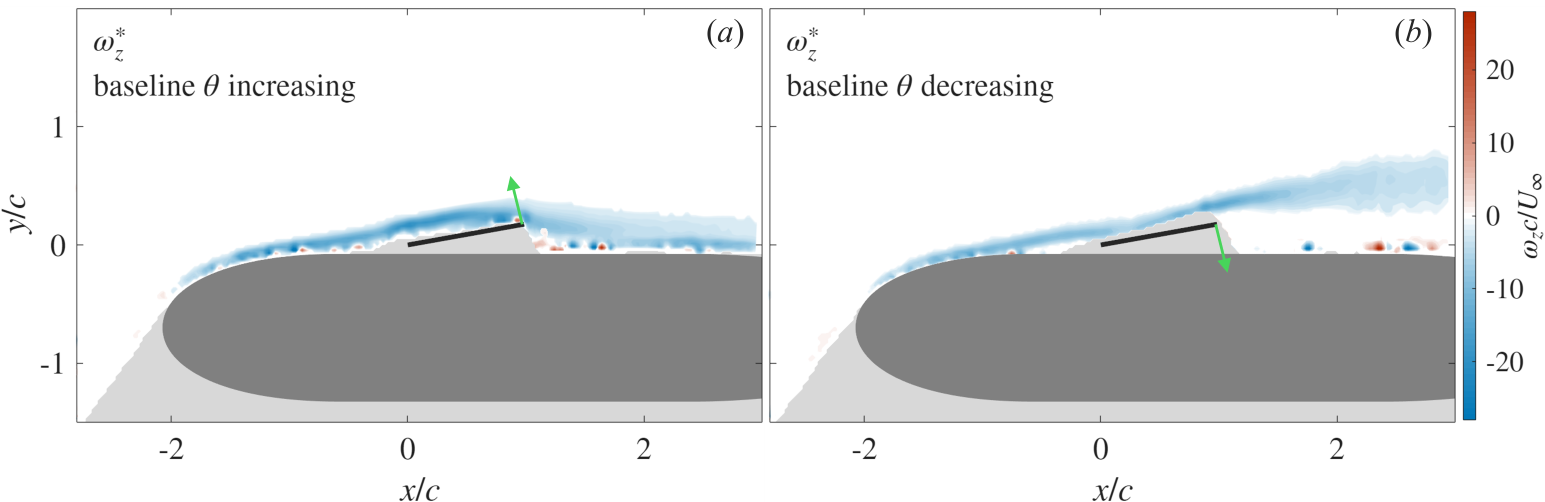}
        \caption{PIV results for the baseline case ($k=0.006$) at $\theta = 10^\circ$:\protect\\ (a) the fin moves away from the body ($\theta$ increasing); (b) the fin moves toward the body ($\theta$ decreasing); the color map shows the normalized vorticity field.}
        \label{fig:results:baseline:piv}
    \end{figure}

    Hysteretic characteristics stand out in the baseline measurements, especially in the lift coefficient curves at low angles. 
    The lift increases as the fin moves away from the body, but then drops sharply, becoming negative for $25 < \theta < 100$.
    As the fin closes, the lift follows a monotonic trend, increases back to zero when the fin is again flat against the body.
    The rise and sudden collapse of the lift force is due to a cambering effect generated by the fin, which breaks down due to flow separation at $\theta > 25$.
    \Cref{fig:results:baseline:piv} shows how the flow structure affects the hysteretic behavior of lift at fin angle $\theta=10^\circ$, where \cref{fig:results:baseline:piv}a shows the time average vorticity field during the flap-up phase as the fin slowly (quasi-steady) moves away from the body, and \cref{fig:results:baseline:piv}b shows the snapshot during the flap-down phase.
    In \cref{fig:results:baseline:piv}a, the shear layer remains continuous along the fin and the body as the fin angle increases (below $25^\circ$) and is cambered near the fin tip, which causes the lift to increase at this stage.
    In contrast, the shear layer is separated during the flap-down phase even at the same fin angle as in \cref{fig:results:baseline:piv}a.
    The separated shear layer does not reattach to the body until the fin closes the gap against the body, which is reflected in the monotonic trends in the lift plot (\cref{fig:results:baseline}a).

    \subsection{Unsteady flapping and vortex evolution}
    \label{sec:results:unsteady}
    For all unsteady cases, the hydrodynamic load measurements are shown by the load \emph{fluctuations}, $\Delta C_L, \Delta C_D$ in which the steady-state loads at $\theta=0^\circ$ are subtracted from the time-varying load coefficients.
    We first discuss the lift (transverse) forces in \cref{sec:results:unsteady:lift}, followed by the drag (streamwise) forces in \cref{sec:results:unsteady:drag}.

        \subsubsection{Lift forces}
        \label{sec:results:unsteady:lift}
\begin{figure}
            \centering
            \includegraphics[width=0.9\textwidth]{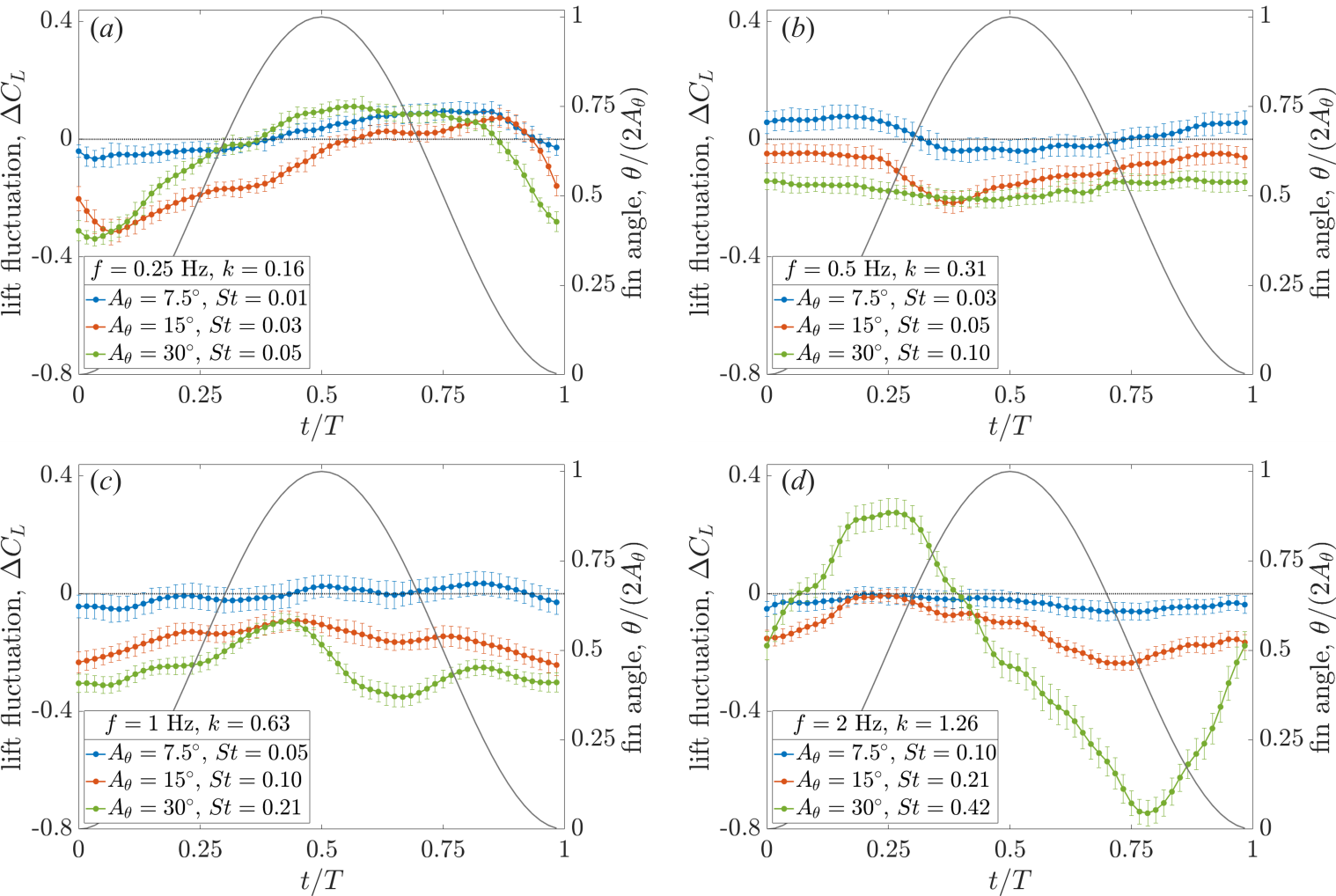}
            \caption{Dynamic lift coefficients fluctuations $\Delta C_L$: (a) $f=0.25\unit{Hz}$, $k=0.16$; (b) $f=0.5\unit{Hz}$, $k=0.31$; (c) $f=1\unit{Hz}$, $k=0.63$; (d) $f=2\unit{Hz}$, $k=1.26$;\protect\\ blue data points: $A_\theta=7.5^\circ$; red data points: $A_\theta=15^\circ$; green data points: $A_\theta=30^\circ$; error bars: $1\sigma$ standard deviation intervals; white solid line: normalized fin angle $\theta/(2A_\theta)$; white dotted line: $\Delta C_L=0$ reference.}
            \label{fig:results:unsteady:CL}
        \end{figure}

        The results are presented in four groups, based on the flapping frequency in \cref{fig:results:unsteady:CL} which shows $\Delta C_L$ vs. the flapping phase, $t/T$, where $T$ is the period of the flapping cycle.
        The gray lines indicate the instantaneous fin angle normalized by the flapping amplitude of the corresponding case.
        The cycle starts at $\theta=0^\circ$ (fin parallel to the body) when $t/T=0$.
        The upstroke phase ends at $\theta=2A_\theta$ when $t/T=0.5$; the downstroke phase brings the fin back to $\theta=0^\circ$ at $t/T=1$.
        
        The mean $\Delta C_L$ value starts to shift to the negative values with respect to the flapping amplitude $A_\theta$ when the flapping frequency reaches $k=0.31$ and above (\cref{fig:results:unsteady:CL}b, c, and d).
        The amplitudes of the lift fluctuation, $A_{C_L}$, also show a growing trend with respect to $A_\theta$ (curves moving from blue to red to green), which translates to larger Strouhal numbers
        when the reduced frequency is fixed. This trend is more evident in the higher-frequency cases.
        
        Most $\Delta C_L$ are negative, which indicates a normal force pushing toward the body.
        The most apparent exception is found in the most unsteady case, $k=1.26$ and $St=0.42$ (green curve in \cref{fig:results:unsteady:CL}d), where $\Delta C_L$ increases to positive values as the fin flaps away from the body at maximum angular velocity.
        This is related to a suction force generated between the fin and the body when the instantaneous angular velocity is high.
        This will be discussed in more detail in the following section in connection with the corresponding unsteady flow fields.
        
        Lastly, we also observe a phase shift across the reduced frequency range. Referenced to the fin angle, the phase shift in the $\Delta C_L$ depends more on the change in the reduced frequency than on that of the Strouhal number.
        This will be examined in more detail \cref{sec:results:scaling} when we discuss scaling.
        
\begin{figure}
            \centering
            \includegraphics[width=0.95\textwidth]{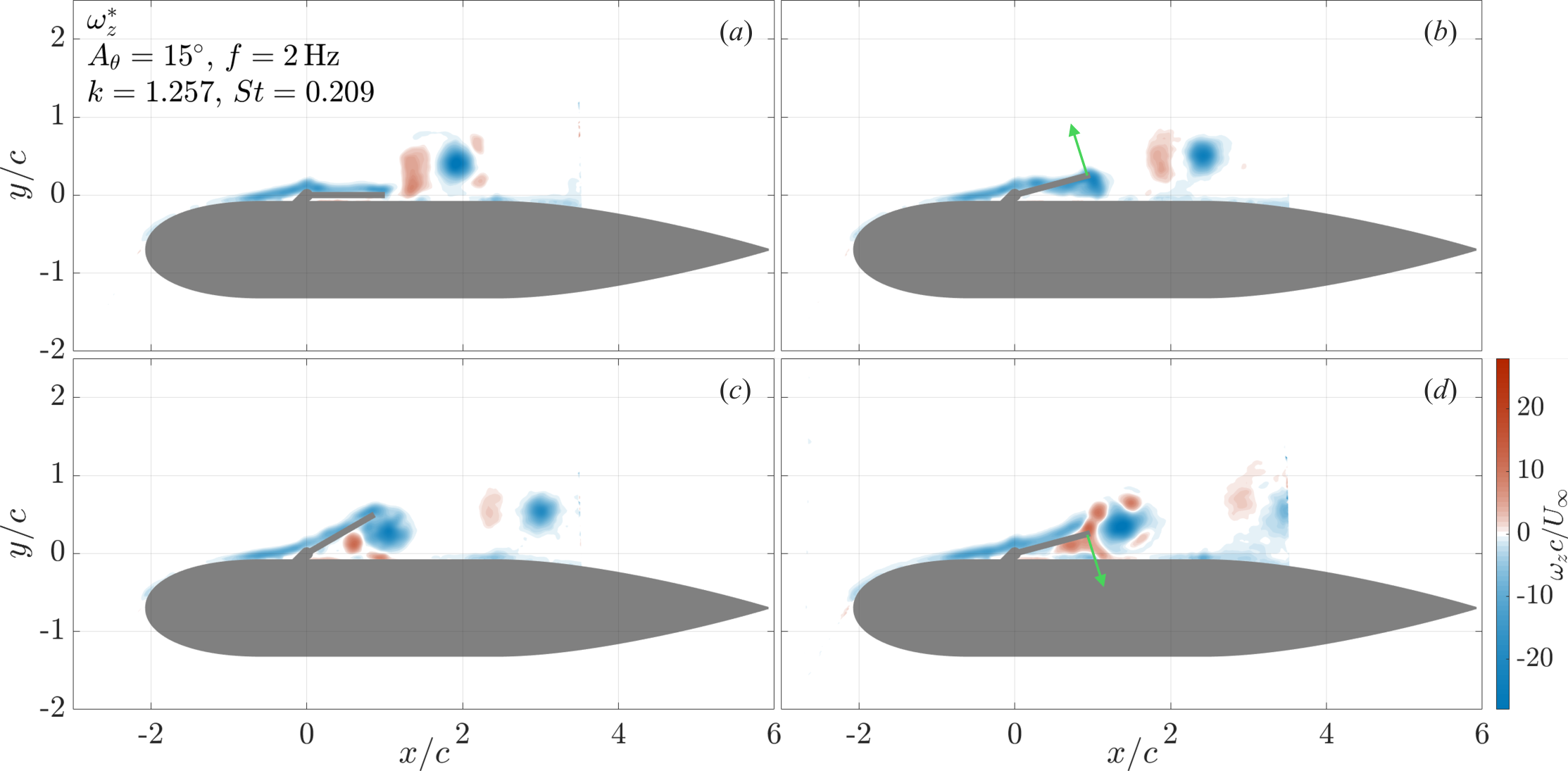}
            \caption{PIV results at $k=1.257$ ($f=2\unit{Hz}$) and $St=0.209$ ($A_\theta=15^\circ$);
            (a) to (d) corresponds to snapshots of flow fields at $t/T=0$, $1/4$, $1/2$, and $3/4$, respectively;\protect\\
            color map: nondimensional vorticity $\omega_z^*$; green arrow: instantaneous flapping direction.}
            \label{fig:results:vortex:A15}
        \end{figure}

\begin{figure}
            \centering
            \includegraphics[width=0.95\textwidth]{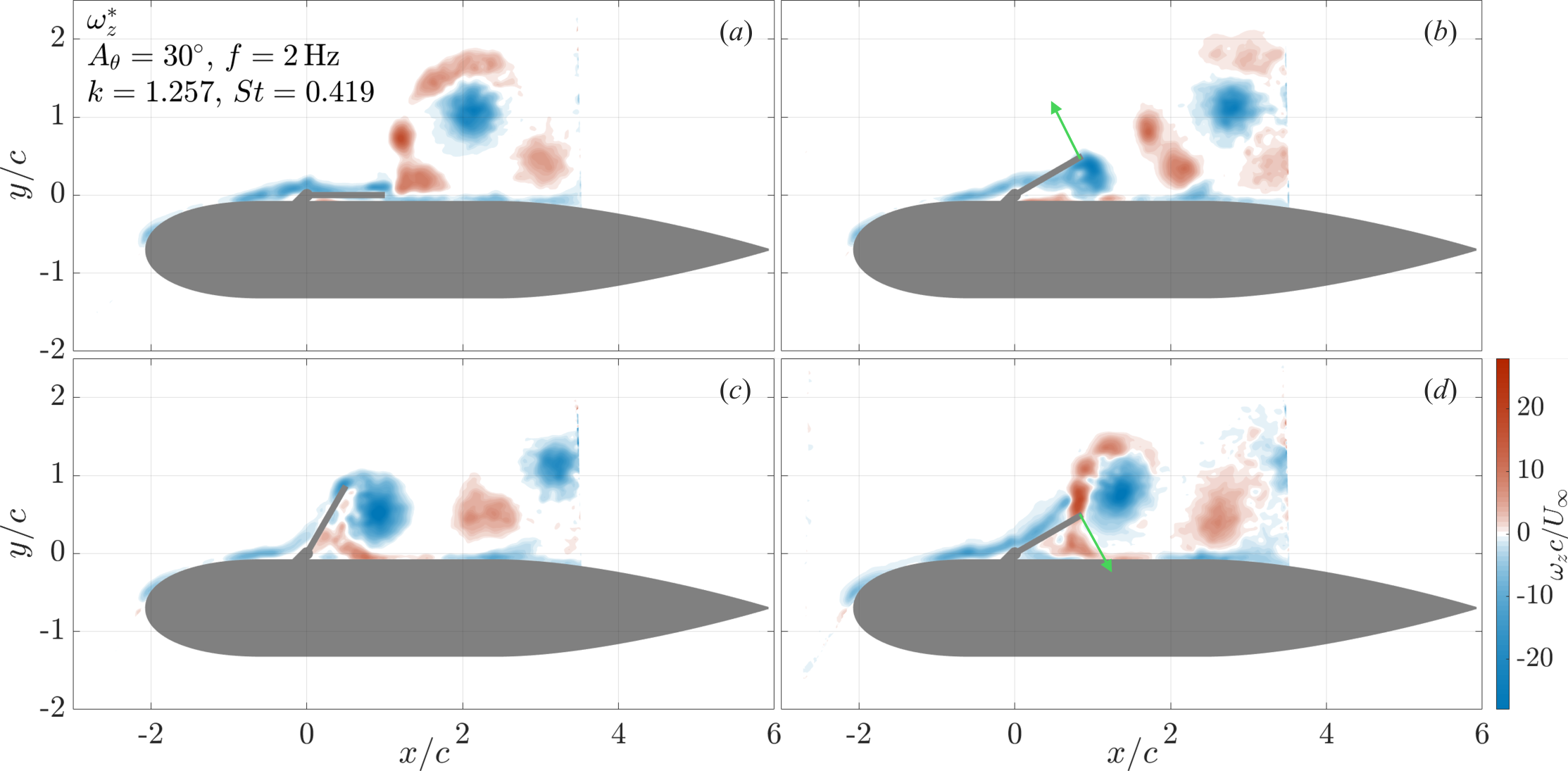}
            \caption{PIV results at $k=1.257$ ($f=2\unit{Hz}$) and $St=0.419$ ($A_\theta=30^\circ$);
            (a) to (d) corresponds to snapshots of flow fields at $t/T=0$, $1/4$, $1/2$, and $3/4$, respectively;\protect\\
            color map: nondimensional vorticity $\omega_z^*$; green arrow: instantaneous flapping direction.}
            \label{fig:results:vortex:A30}
        \end{figure}

        Selected out-of-plane vorticity fields, $\omega_z$,  for $k=1.26$ ($f=2\unit{Hz}$) are shown in \cref{fig:results:vortex:A7.5,fig:results:vortex:A15,fig:results:vortex:A30}.
        We show the case of $k=1.26$ (2 Hz) as an example case chosen to illustrate how the lift fluctuations, $\Delta C_L$, evolve with the Strouhal number (videos of the PIV results are included in the supplemental material).
        
        While the frequency is fixed at $k=1.26$, each figure shows the velocity fields for a different flapping amplitude, $A_\theta = 7.5,15$ and 30$^\circ$ corresponding to Strouhal numbers,  $St=0.10$, $0.21$, and $0.42$ (corresponding to the force plots in \cref{fig:results:unsteady:CL}d).
        Each figure shows four snapshots of the flow fields at $t/T=0$, $1/4$, $1/2$, and $3/4$ phases of the flapping cycle.
        The green arrows indicate the direction and magnitude of the fin motion (there is no arrow shown in sub-figures a and c since the instantaneous flapping rate at minimum ($t/T=0$) and maximum ($t/T=1/2$) positions is zero).
        
        In all cases, the vorticity from the leading-edge boundary layer is fed to and accumulated at the fin tip as it flaps up, thus generating a clockwise-rotating main vortex, which is shown by the negative (blue) color map.
        
        This vortex starts to shed as the fin approaches its maximum position (sub-figures b to c in \cref{fig:results:vortex:A7.5,fig:results:vortex:A15,fig:results:vortex:A30}) where the fin decelerates to a flapping rate at which the vorticity feeding the main vortex is not sufficient to keep it attached to the fin tip.
        After reaching the maximum flapping angle at $t/T=1/2$ (sub-figures c), the main vortex separates and is then advected downstream by the mean flow, moving away from the body.  
        
        During the main vortex formation and shedding process, secondary processes also occur depending on the Strouhal number.
        One is the generation of counter-rotating secondary vortices at the fin tip during the downstroke phase at all Strouhal numbers, which can be observed from the positive (red) color map in sub-figures d.
        The other secondary process is a suction effect at the corner between the fin and the body during the upstroke phase at a higher Strouhal number (\cref{fig:results:vortex:A30}c).    
        The last secondary process is an accelerated flow, appearing as a ``jet'', generated between the fin and the body during the downstroke phase (see \cref{fig:results:vortex:A15,fig:results:vortex:A30}d to a).
        This secondary process relates more to the drag fluctuation behavior, which will be discussed more in \cref{sec:results:unsteady:drag}.
        
        The secondary tip vortices are induced by the opposite fin flapping rate and acceleration during the downstroke phase.
        The vorticity is produced on the lower surface of the fin by the fin motion and on the body surface by the reverse shear induced by the rotation of the main vortex.
        Distinct from the main vortex formation process, there is not enough vorticity on the lower side of the fin that can be fed to the tip at a sufficient rate to sustain the formation of a larger vortex.
        Therefore, instead of a counter-rotating vortex with a comparable size to the main vortex, several smaller vortices are shed at the fin tip; the number and strength of these secondary vortices are influenced by the Strouhal number.
        A larger Strouhal number results in stronger and more secondary vortices, whereas fewer and weaker secondary vortices are generated in small Strouhal number cases.
        In some of the smallest Strouhal number cases, the secondary vortices are too weak to be visually observed in the PIV measurements.
        All of the secondary vortices demonstrate a clockwise orbiting trend around the main vortex, as the strength and size of the main vortex dominate during the convection (see supplementary videos).
        
        The suction effect during the upstroke phase (\cref{fig:results:vortex:A30}c) is a result of fluid continuity, where the flow needs to fill the space that is created as the fin departs from the body surface.
        This occurs when the local flow stagnates or moves against the mean flow near the corner area, which also increases the rotation of the main vortex and produces opposite-sign vorticity through wall shear, as mentioned above.
        Intuitively, and as the PIV results confirm, this suction effect is more apparent at high $St$ since the fin acceleration is larger in these cases, and the rate of space being created is faster, so the local flow needs to move backward to fill that space more rapidly.
        \begin{figure}
            \centering
            \includegraphics[width=0.95\textwidth]{fig/results_vortex_piv_A7.5_f2Hz.png}
            \caption{PIV results at $k=1.257$ ($f=2\unit{Hz}$) and $St=0.105$ ($A_\theta=7.5^\circ$); 
            (a) to (d) corresponds to snapshots of flow fields at $t/T=0$, $1/4$, $1/2$, and $3/4$, respectively;\protect\\
            color map: nondimensional vorticity $\omega_z^*$; green arrow: instantaneous flapping direction.}
            \label{fig:results:vortex:A7.5}
        \end{figure}

        \subsubsection{Drag forces}
        \label{sec:results:unsteady:drag}
\begin{figure}
            \centering
            \includegraphics[width=0.9\textwidth]{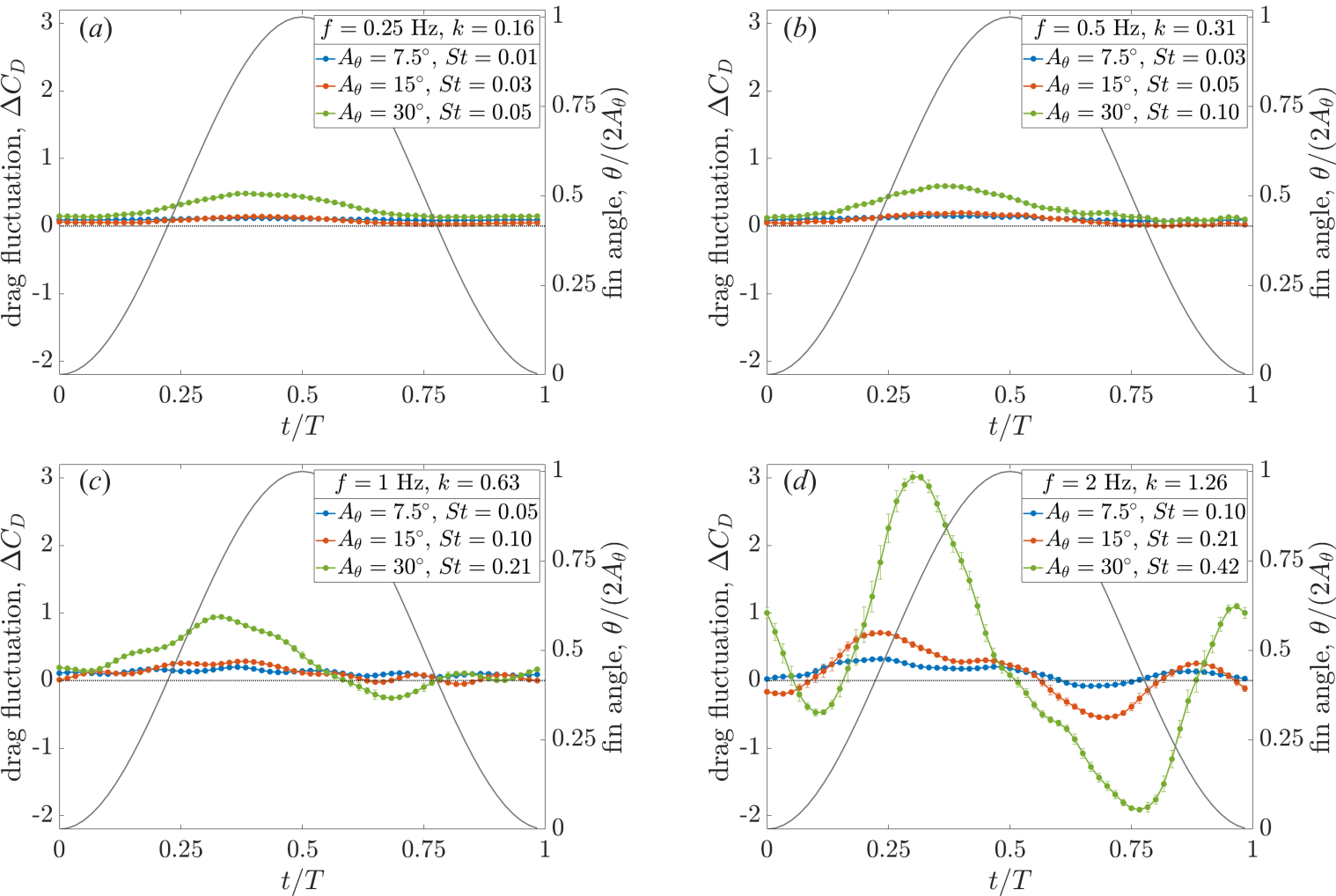}
            \caption{Dynamic drag coefficients fluctuations $\Delta C_D$: (a) $f=0.25\unit{Hz}$, $k=0.16$; (b) $f=0.5\unit{Hz}$, $k=0.31$; (c) $f=1\unit{Hz}$, $k=0.63$; (d) $f=2\unit{Hz}$, $k=1.26$;\protect\\ blue data points: $A_\theta=7.5^\circ$; red data points: $A=15_\theta^\circ$; green data points: $A_\theta=30^\circ$; error bars: $1\sigma$ standard deviation intervals; white solid line: normalized fin angle $\theta/(2A_\theta)$; white dotted line: $\Delta C_D=0$ reference.}
            \label{fig:results:unsteady:CD}
        \end{figure}

        The drag fluctuations, $\Delta C_D$, shown in \cref{fig:results:unsteady:CD}, demonstrate a monotonic trend in amplitude with respect to both the flapping frequency and flapping amplitude. Using the fin angle (gray lines) as the reference, all positive $\Delta C_D$ peaks occur before the maximum angle is reached at $t/T=1/2$.
        Referring back to the flow fields (\cref{fig:results:vortex:A7.5,fig:results:vortex:A15,fig:results:vortex:A30}), there are no obvious, large-scale flow structures that correlate with the drag rise.
        This suggests that the profile drag due to the geometry change and the added mass effect of the fin motion may be the cause of most of the $\Delta C_D$ increment.
        In addition, the momentum effect of the backward motion of the local flow under the fin as the fin flaps upward (sub-figures b \& c of \cref{fig:results:vortex:A7.5,fig:results:vortex:A15,fig:results:vortex:A30}) can also contribute to the increase in $\Delta C_D$ in the higher flapping frequency cases.
        
        In some high $St$ cases, we observe \emph{thrust} generation, $\Delta C_D < 0$, which deserves further attention.
        As the magnitude of the drag fluctuation grows with the Strouhal number, $\Delta C_D$ starts to produce negative values when $St\ge 0.209$ (green curve in \cref{fig:results:unsteady:CD}c and the red and green curves in \cref{fig:results:unsteady:CD}d), which occurs between $t/T=0.55$ and $0.77$ depending on the specific case.
        
\begin{figure}
            \centering
            \includegraphics[width=0.5\textwidth]{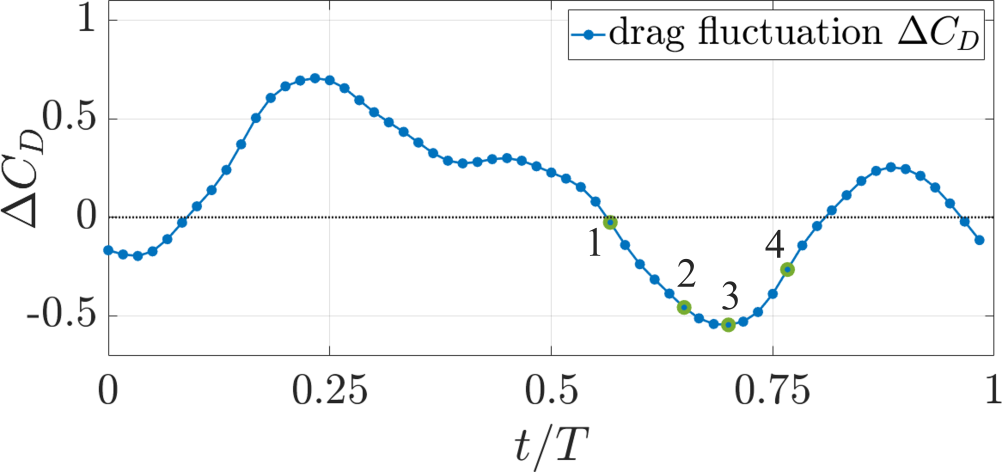}
            \caption{Drag coefficient variation at $k=1.26$ ($f=2\unit{Hz}$) and $St=0.209$ ($A_\theta=15^\circ$): the dotted white line is the $\Delta C_D=0$ reference; numbers mark out the data points of interest, where the fin flaps down, and thrust is generated (negative $\Delta C_D$).}
            \label{fig:results:vortex:drag}
        \end{figure}

        As an example, we can examine a non-extreme case at $k=1.26$ ($f=2\unit{Hz}$) and $St=0.209$ ($A_\theta=15^\circ$), (\cref{fig:results:vortex:A15} and \cref{fig:results:unsteady:CD}d, red line), where the thrust (negative $\Delta C_D$) is still generated during the downstroke phase.
        \Cref{fig:results:vortex:drag} reproduces this curve for convenience, and timestamps during the thrust-generating phase from $t/T=0.58$ to $0.8$ are labeled in green on the $\Delta C_D$ curve.
        The corresponding horizontal velocity ($u^*$) fields are shown in \cref{fig:results:vortex:drag:piv} (also see the supplementary video), of which the color map shows the backward velocity in deep blue, forward velocity lower than freestream velocity in light blue, and forward velocity greater than the freestream velocity in red.
        The counter-direction velocity blobs are a result of the rotation of the main vortex.
        There is a part of the flow that is entrained into the corner between the fin and body during the extension phase of the fin, which appears as a velocity blob at the freestream velocity in the horizontal velocity plots.
        As the fin accelerates toward the body (\cref{fig:results:vortex:drag:piv}b-d), the flow at the corner is pushed out downstream as a local jet generated by the closing motion of the fin.
        The momentum carried by the jet also pushes the main vortex downstream, thus appearing as a thrust in hydrodynamic load measurement.
        
        \begin{figure}
            \centering
            \includegraphics[width=0.88\textwidth]{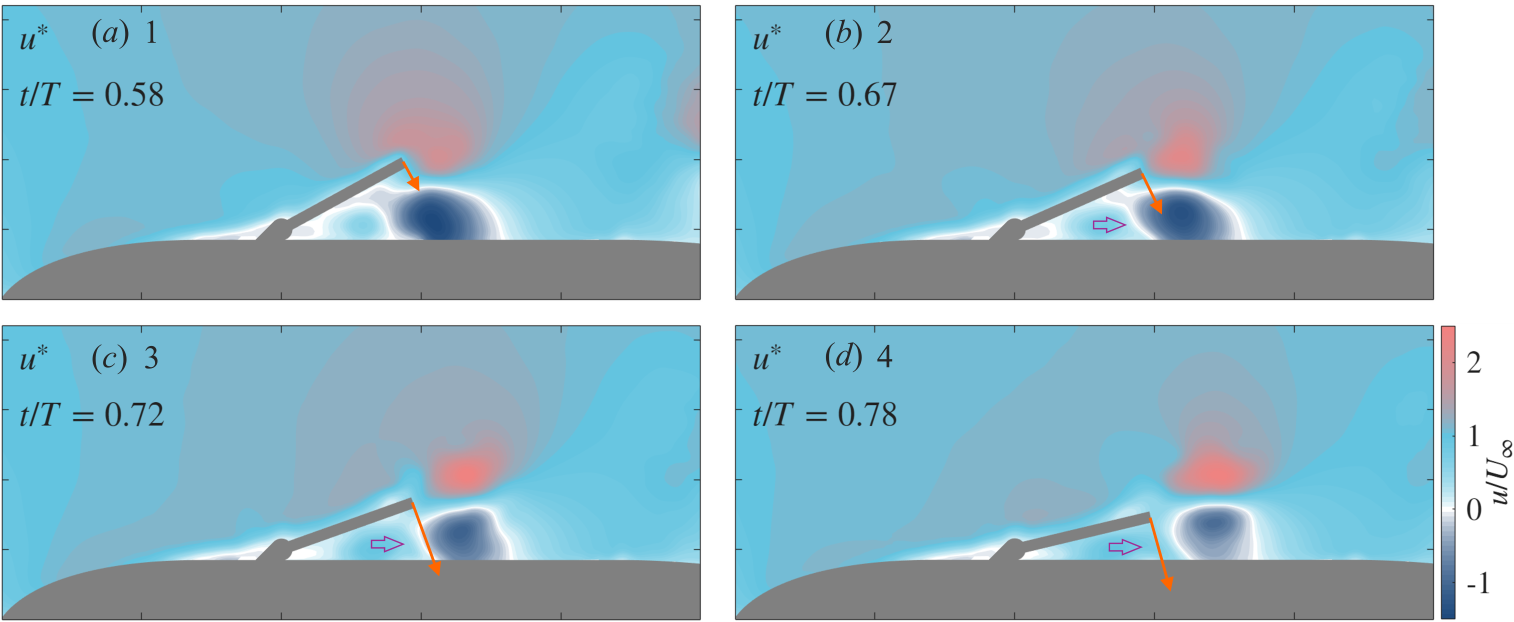}
            \caption{Flow fields of interested data points in drag fluctuation; color map: nondimensional horizontal velocity, $u^*$; numbers correspond to the points of interest shown in \cref{fig:results:vortex:drag}; yellow arrow: direction and amplitude of flapping rate; the fin is flapping downward in all snapshots.}
            \label{fig:results:vortex:drag:piv}
        \end{figure}

    \subsection{Data-driven scaling}
    \label{sec:results:scaling}
    For either lift and drag fluctuations, the growth of the fluctuation amplitude ($A_{C_L}$ or $A_{C_D}$) is not linearly correlated with any of the controlled parameters such as reduced frequency, $k$, Strouhal number, $St$, or flapping amplitude, $A_\theta$.
    However, from \cref{fig:results:unsteady:CL,fig:results:unsteady:CD}, one can intuitively suspect that there could be scaling relations between the fluctuation amplitudes and reduced frequencies and/or Strouhal numbers, as they commonly appear in previous scaling studies \citep{Triantafyllou1993,quinn2014pitching,dewey2013,floryan2017}.
    Instead of theoretically deriving the possible scaling relations for such a complex fin-body configuration, we try to scale the hydrodynamic loads through a data-driven approach, taking advantage of the relatively large data set. 
    
    The Sparse Identification of Nonlinear Dynamics (SINDy) algorithm \citep{brunton2016} is an efficient tool to identify dynamic models or governing equations from measurements.
    We use a modified formulation of SINDy \citep{he2023} to identify the scaling relations from the experimental data set, since the scaling relations are not treated as dynamic systems.
    
    The input-output relation for the SINDy algorithm can be written as
    \begin{equation}
        \bm{Y} = \bm{\varTheta}(\bm{X})\bm{\varXi},
        \label{eq:results:scaling:sindy}
    \end{equation}
    where $\bm{Y}$ is the training data set, $\bm{\varTheta}(\bm{X})$ is a matrix that serves as a library of candidates of scaling terms, $\bm{X}$ is the state variable vector, and $\bm{\varXi}$ is the coefficient vector to be identified.
    In the original form of SINDy, $\bm{Y}$ is the time derivative of the state variable $\bm{X}$, and $\bm{\varTheta}$ contains the state variable as shown.
    Here, however, we do not treat the scaling relation as a time-varying system as mentioned; therefore, a non-dynamic formulation is used, where
    \begin{equation}
        \bm{Y} = \bm{X} = \bm{A_{C_L}}\; \text{or}\; \bm{A_{C_D}}
    \end{equation}
    and, initially,
    \begin{equation}
        \bm{\varTheta} = [St\quad k\quad A^*\quad St^2\quad kSt\quad A^*St\quad St^2k]^\mathrm{T},
    \end{equation}
    which are the candidates of the possible scaling terms based on an initial guess of possible first- and second-order terms and an additional higher-order term $St^2k$ as suggested by \citet{floryan2017}.
    Note that $A^*$ is the flapping amplitude in radians.
    
\begin{figure}
        \centering
        \includegraphics[width=0.5\textwidth]{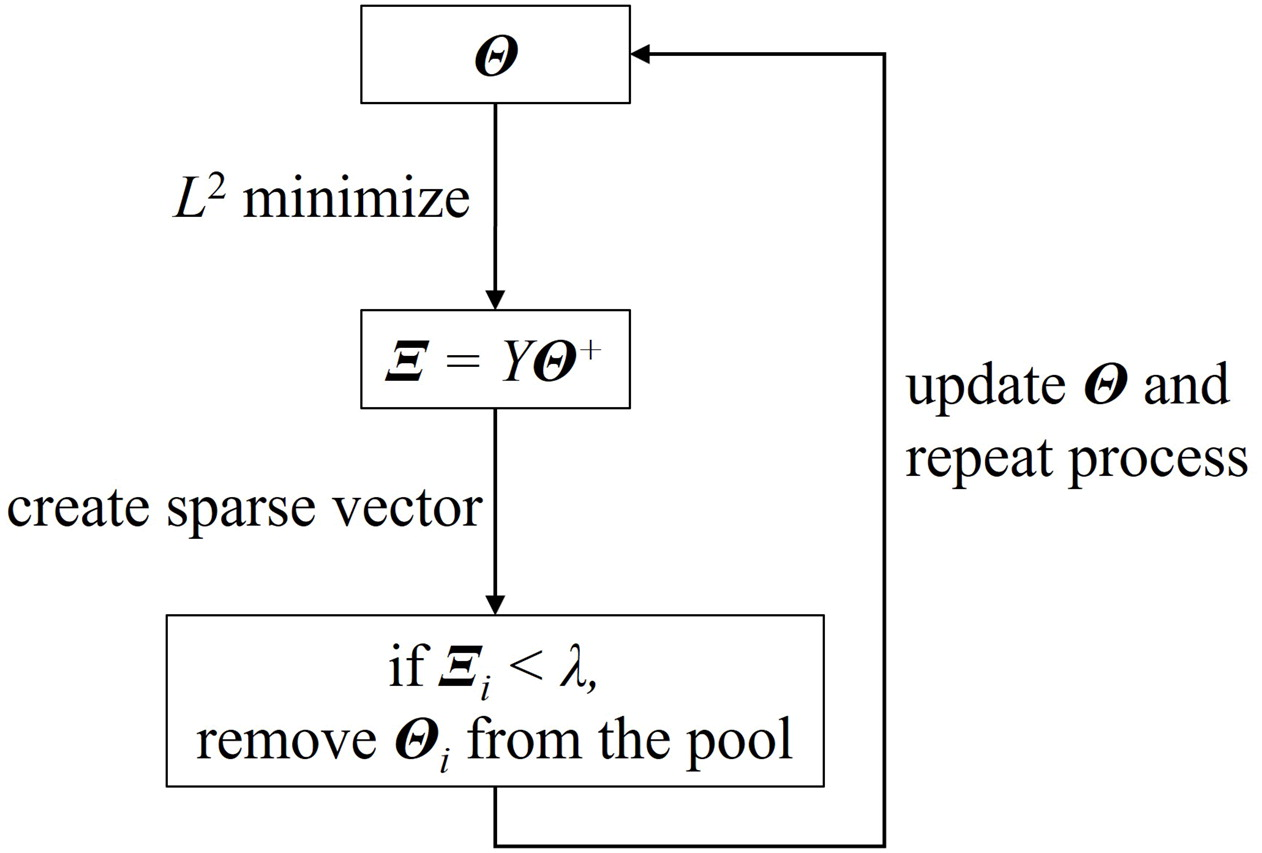}
        \caption{Flowchart of SINDy algorithm \citep{he2019:sindy}.}
        \label{fig:results:scaling:sindyflowchart}
    \end{figure}

    As shown in \cref{fig:results:scaling:sindyflowchart}, the goal is to find the $\bm{\varXi}$ vector using sparse regression techniques.
    An L2-norm minimization is performed on \cref{eq:results:scaling:sindy} to generate the coefficients in $\bm{\varXi}$ in each iteration.
    If there is any $\varXi_i<\lambda$, then the corresponding $\varTheta_i$ is excluded from the candidate library and $\bm{\varTheta}$ is updated.
    This process is repeated until all elements of $\bm{\varXi}$ satisfy $\varXi_i>\lambda$.
    The sparsity parameter, $\lambda$, controls how many terms will be included in the final identification results.
    The larger $\lambda$ is, the fewer candidates will be kept, and the resulting system is more ``sparse''.
    Small $\lambda$ values include more terms, improving the fitting results of the training data.
    This, however, is not necessarily a better system as it can result in over-fitting the data and losing physical meaning and robustness.
    
\begin{figure}
        \centering
        \includegraphics[width=0.82\textwidth]{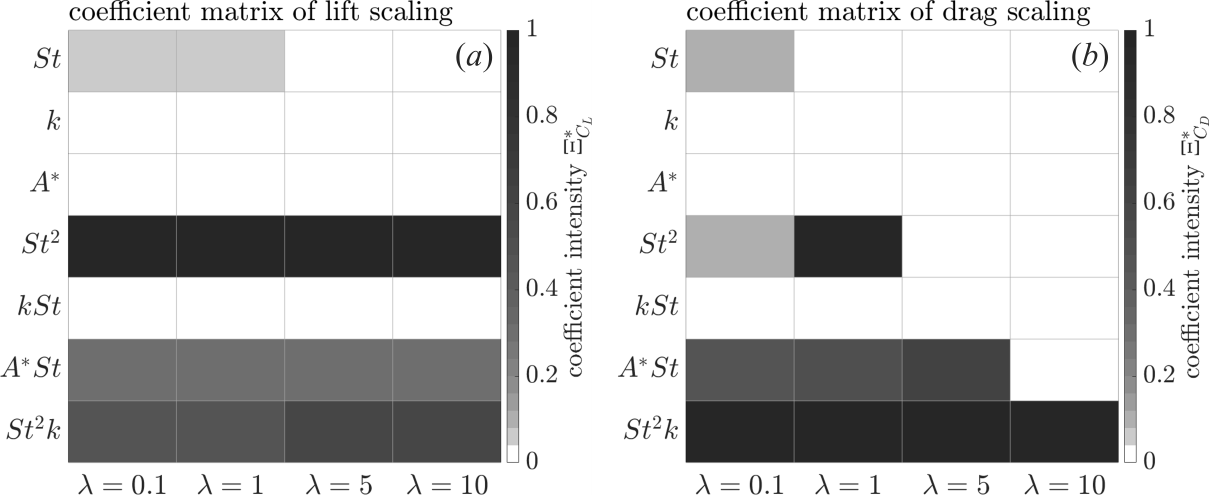}
        \caption{SINDy coefficients identification results: (a) lift scaling; (b) drag scaling; the ordinate is the candidate terms, the abscissa is the sparsity parameter $\lambda$; each column is a set of identified coefficients corresponding to the specific $\lambda$; color map indicates the coefficient $\varXi$ normalized by the maximum value of each set.}
        \label{fig:results:scaling:sindy}
    \end{figure}

    The SINDy identification results for lift scaling and drag scaling are visualized in terms of coefficient intensity in \cref{fig:results:scaling:sindy}a and b, respectively.
    Each column represents a coefficient vector, $\bm{\varXi}$, identified using the corresponding $\lambda$ on the abscissa.
    The elements of the coefficient vectors are normalized by their maximum, such that the relative weight, $\varXi^*$, of each candidate term can be shown by the color map.
    For example, in the identification results for lift scaling, when $\lambda=5$, candidate terms, $St$, $k$, $A^*$, and $kSt$, have coefficients of zeros in the final $\bm{\varTheta_{C_L}}$ vector, which means that they are dropped from from the regression iterations because their coefficients are below the threshold defined by this value of $\lambda$.
    The remaining terms $St^2$, $A^*St$, and $St^2k$ are the scaling terms to be used to scale all lift coefficient data with their corresponding $\varXi_{C_L}$ values.
    
    In the case of lift scaling, $St^2$ is the most influential term, and most nonlinear terms of $St$ play significant roles, as indicated by the selection from SINDy iterations.
    The identification of drag scaling is interpreted in the same way. In addition, the candidate selection results are more sensitive to the sparsity parameter as shown in \cref{fig:results:scaling:sindy}b.
    The number of identified scaling terms in the $\bm{\varTheta_{C_D}}$ vector decreases from 4 to 1 as $\lambda$ changes from $0.1$ to $10$.
    
    As mentioned in the SINDy formulation, a reasonable sparsity parameter is desired to prevent the system from being either inaccurate (too sparse) or not meaningful (overfitting).
    Since they are using the same sets of training data, an intermediate sparsity parameter, $\lambda=1$, is chosen for both lift and drag scaling to be consistent. With this choice, we obtain the scaled lift coefficient, $\Delta C_L^*$, and drag coefficient, $\Delta C_D^*$, as
    \begin{equation}
        \Delta C_L^* = \frac{\Delta C_L}{\bm{a_{C_L}} [St\quad A^*\quad St^2\quad kSt\quad A^*St\quad St^2k]^\mathrm{T}} , 
        \label{eq:results:scaling:CL}
    \end{equation}
    and
    \begin{equation}
        \Delta C_D^* = \frac{\Delta C_D}{\bm{a_{C_D}} [St^2\quad A^*St\quad St^2k]^\mathrm{T}},
        \label{eq:results:scaling:CD}
    \end{equation}
    where vectors $\bm{a_{C_L}}$ and $\bm{a_{C_D}}$ contain the coefficients of their corresponding terms,
    \begin{equation}
        \bm{a_{C_L}} = [-6.66,\,2.54,\,105.02,\,2.67,\,-29.62,\,-48.91]
        \end{equation}
    and
    \begin{equation}
        \bm{a_{C_D}} = [-33.78,\,16.35,\,33.32].
    \end{equation}
        
\begin{figure}
        \centering
        \includegraphics[width=0.9\textwidth]{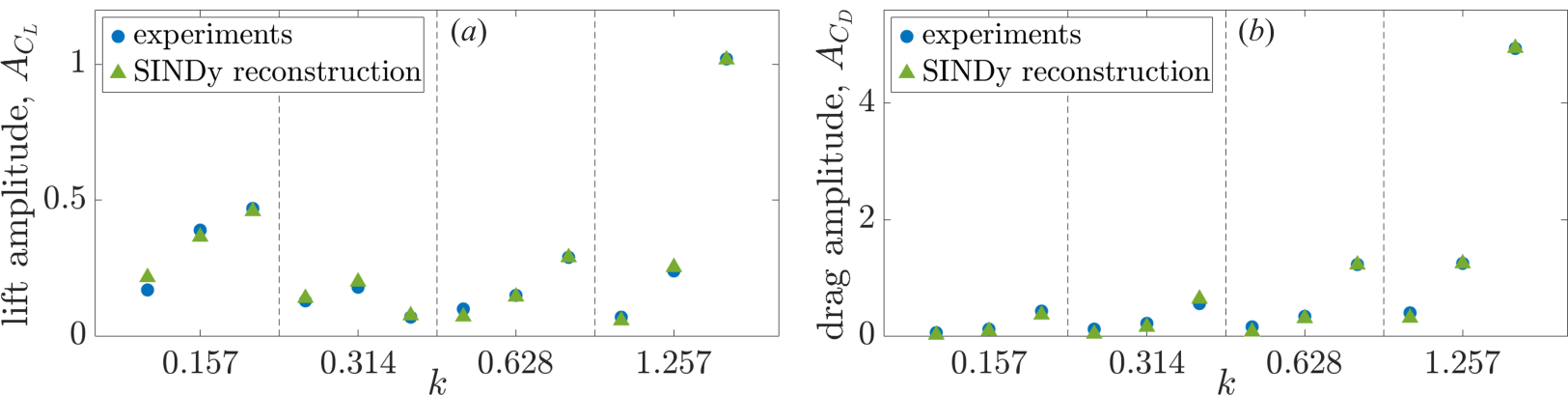}
        \caption{Amplitude scaling results: (a) lift amplitudes; (b) drag amplitudes; vertical dotted lines separate the data into groups based on the reduced frequencies; blue circles are the experimental results (training data); green triangles are the reconstruction results using identified terms and their coefficients (validation).}
        \label{fig:results:scaling:amp}
    \end{figure}

    We examine the accuracy of the scaling by using \cref{eq:results:scaling:CL,eq:results:scaling:CD} to recover the fluctuation amplitudes and compare them with the measurements.
    The comparisons are shown in \cref{fig:results:scaling:amp}, where the data are grouped by their reduced frequencies.
    The data points in the same $k$ column have the same reduced frequency, and the Strouhal number, $St$, increases from left to right within the same $k$ column.
    The errors of the scalings are taken as the ratios between the absolute errors and the corresponding amplitudes of the experimental results, where the error $Err=|A_\theta-A_\mathrm{scaling}|/A_\theta$.
    Overall, the lift scaling is more accurate relative to the drag scaling. Some of the cases have relatively high errors due to the small amplitudes of the unscaled loads in those cases, for instance, the drag scaling at low $k$.
    The accuracy of both scalings increases at higher reduced frequencies, especially when the performances of lift and drag scalings are comparable at the highest-tested $k$.
    At a fixed $k$, the accuracy of the scalings improves with increasing $St$. This is expected as the amplitude of load variations demonstrates a more monotonic growth at higher $k$ and $St$, especially for the lift variations (\cref{fig:results:unsteady:CL}), and it is easier to identify terms and their coefficients from such data sets, so the final result favors these cases.
    The $St$ effect on the accuracy, in addition, helps improve the scaling performance by creating a larger signal-to-noise ratio in high-amplitude cases.
        
    \begin{figure}
        \centering
        \includegraphics[width=0.5\textwidth]{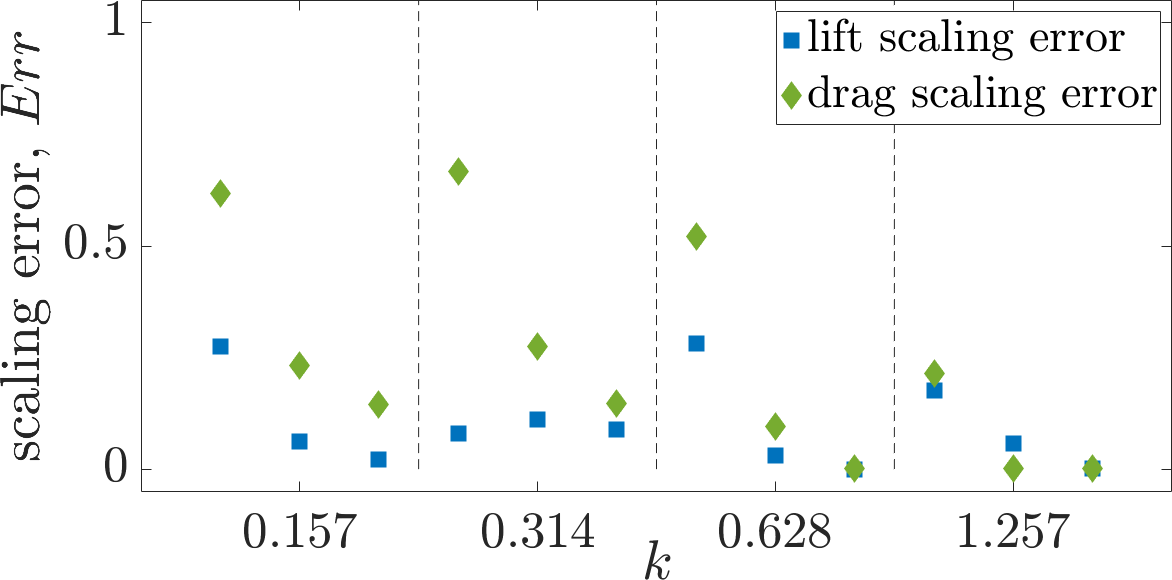}
        \caption{Scaling errors: ratio of absolute scaling errors to their corresponding amplitudes measured in experiments; blue squares: lift scaling errors; green diamonds: drag scaling errors.}
        \label{fig:results:scaling:error}
    \end{figure}
    
\begin{figure}
        \centering
        \includegraphics[width=0.9\textwidth]{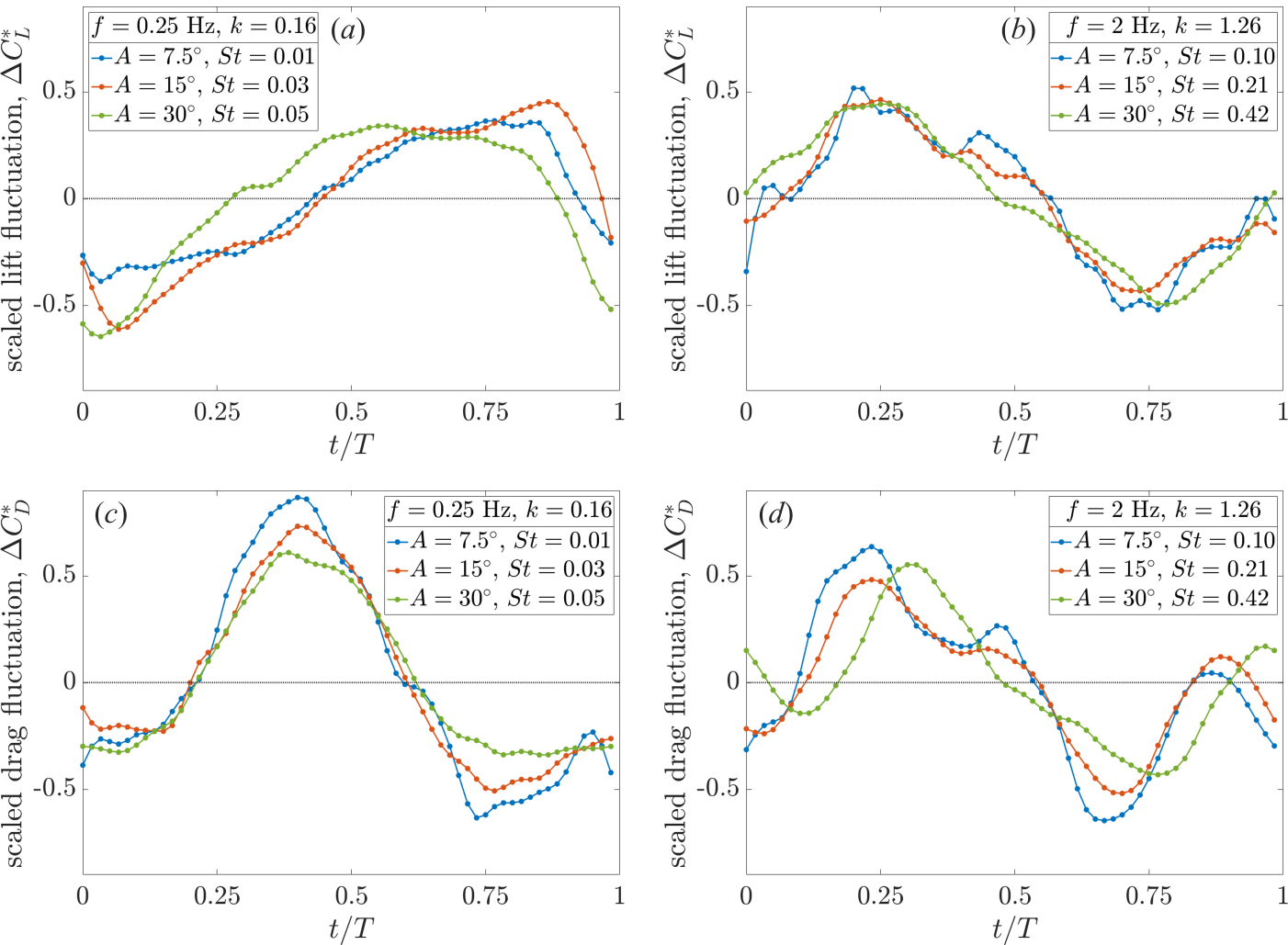}
        \caption{Scaled load fluctuations: (a) and (b) are scaled lift fluctuations, $\Delta C_L^*$, at $k=0.157$ ($f=0.25\unit{Hz}$) and $k=1.257$ ($2\unit{Hz}$); (c) and (d) are scaled drag fluctuations, $\Delta C_D^*$, at $k=0.157$ ($f=0.25\unit{Hz}$) and $k=1.257$ ($2\unit{Hz}$); blue data points: $A=7.5^\circ$; red data points: $A=15^\circ$; green data points: $A=30^\circ$; white dotted line: zero-value reference.}
        \label{fig:results:scaling:vld}
    \end{figure}

\begin{figure}
        \centering
        \includegraphics[width=0.9\textwidth]{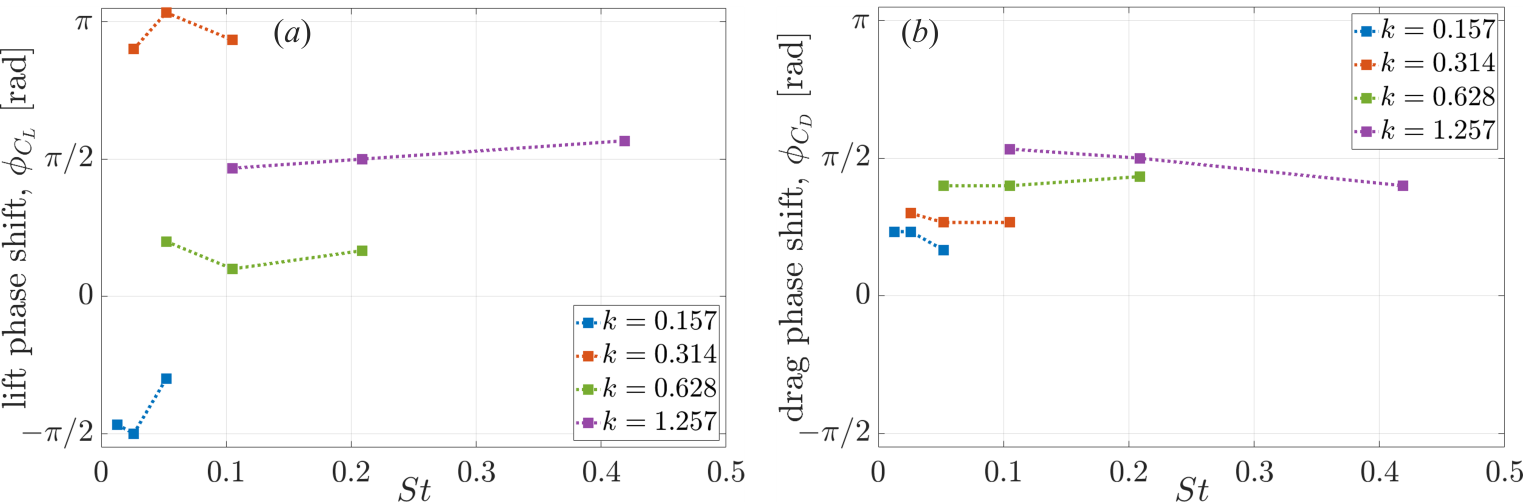}
        \caption{Phase lag between the scaled hydrodynamic loads and the fin flapping angle: (a) phase lag $\phi_{C_L}$ between lift coefficients $\Delta C_L^*$ and fin angle $\theta$; (b) phase lag $\phi_{C_D}$ between drag coefficients $\Delta C_D^*$ and fin angle $\theta$; blue data points: $k=0.157$; red data points: $k=0.314$; green data points: $k=0.628$; purple data points: $k=1.257$.}
        \label{fig:results:scaling:phaseshift}
    \end{figure}

    \Cref{fig:results:scaling:vld} shows some example results of the scaled time-series lift and drag fluctuations (mean removed) as a validation of the SINDy-scaling method.
    At a fixed reduced flapping frequency, $k$, the scaling is mainly determined by $St$ appearing in the nonlinear terms, as indicated by the identified coefficients.
    We notice phase shifts across the cases from low to high $St$ and $k$. Using the fin angle $\theta$ as the reference, the phase shifts of the scaled lift ($\phi_{C_L}$) and drag ($\phi_{C_D}$) coefficients of all test cases are summarized in \cref{fig:results:scaling:phaseshift}.
    The phase shift data are grouped based on their reduced frequencies and plotted with respect to the specific Strouhal numbers. From the results of the test cases, the reduced frequency has a greater influence on the phase shift than the Strouhal number.
    At fixed reduced frequencies, the maximum difference among the phase shifts is $\Delta\phi_{C_L}=0.17\pi$ found at $k=0.157$ for the lift coefficient and $\Delta\phi_{C_D}=0.13\pi$ at $k=1.257$ for the drag coefficient.
    The phase shift shows a more apparent dependency on the reduced frequency when compared across the tested $k$ values.
    Although the direction of the phase shift of the lift coefficient is nonmonotonic with respect to $k$, the ``mean'' phase shift is decided by the specific $k$ as demonstrated by the four groups of data points in \cref{fig:results:scaling:phaseshift}a.
    For the drag coefficient, the phase shifts are not only separated into four groups by the reduced frequency, but they also lead the reference (fin flapping angle) more as the reduced frequency increases, which shows a more uniform trend than that of the lift coefficient phase shift. 
    
        
\section{Conclusions}
\label{sec:conclusions}
The hydrodynamic lift and drag and the associated vortex evolution of a fin-body configuration are examined in dynamic fin-flapping cases.
Both the lift and drag fluctuations during a flapping cycle are significantly influenced by the reduced frequency and Strouhal number, which manifest as time-series variations and amplitudes.
The flow field mechanism behind the fluctuations is studied through PIV measurements, which reveal the boundary layer and vortical flow behaviors at varying flapping frequencies and amplitudes.
From the PIV results, the formation, growth, and shedding characteristics of the fin tip vortices, including a main vortex and orbiting secondary vortices, are revealed.
However, additional factors, such as added mass, may be needed to fully describe the lift behaviors. 

The thrust generation mechanism in the drag variation at higher Strouhal numbers is analyzed in detail through the PIV results.
The momentum effect of a local jet produced by the downward fin acceleration is the major contributor to the thrust observed in force measurements.

Starting from the visual correlation between the unsteadiness (reduced frequency, $k$, and Strouhal number, $St$) and the fluctuation amplitude of the hydrodynamic loads, scaling relations are sought for both the lift and drag.
A data-driven approach using the SINDy algorithm is employed to mitigate the complexity introduced by the scaling effects of the fin-body configuration.
The identification results show that the quadratic and a few nonlinear combinations of the Strouhal number are the most influential terms in both lift and drag scaling.
In particular, all selected terms for drag scaling are nonlinear terms of the Strouhal number.
The validations of the scaling using the time-series data show that the scalings are robust across the test cases. The performance of the scaling is affected by both $k$ and $St$, where the results show that the scaling prefers monotonic groups of data points and larger amplitudes due to the fundamental mechanism of the SINDy regression.
This data-driven approach provides an alternative perspective for scaling or nondimensional studies of unsteady loads acting on complex configurations.
It can also be extended to other characteristics contained in the data, such as the relation among the phase shift, reduced frequency, and Strouhal number.

\backsection[Acknowledgements]{The authors would like to thank the support from the Office of Naval Research with program officer Dr. Teresa McMullen.}

\backsection[Funding]{Office of Naval Research Award N00014-21-1-2816}

\backsection[Declaration of interests]{The authors report no conflict of interest.}




\bibliographystyle{jfm}
\bibliography{jfm}

\end{document}